\documentclass[11pt]{amsart}
\usepackage[margin=1in]{geometry}
\usepackage{amssymb,xcolor,colortbl,diagbox,comment,graphicx,mathpazo,cite}
\usepackage{MnSymbol}
\usepackage{color,float,fullpage,mathrsfs,mathtools}
\usepackage{enumerate,enumitem}
\usepackage[colorlinks=true, pdfstartview=FitV, linkcolor=blue, citecolor=blue, urlcolor=blue]{hyperref}
\usepackage{multirow,array}
\usepackage{makecell}

\def\sech{\mathrm{sech}}
\newtheorem{theorem}{Theorem}

\numberwithin{equation}{section}

\title{Investigation of shallow water waves near the coast or in lake environments via the KdV-Calogero-Bogoyavlenskii-Schiff equation}

\author{Peng-Fei Han}
\address[P. F. Han]{}
\email{hanpf1995@163.com; 448024030@qq.com}
\author{Yi Zhang$^{*}$}
\address[Corresponding author: Y. Zhang]{Department of Mathematics\\ Zhejiang Normal University\\ Jinhua 321004\\ People's Republic of China.}
\email{zy2836@163.com}

\thanks{$*$ Corresponding author. E-mail address: zy2836@163.com(Yi Zhang)}

\dedicatory{Department of Mathematics, Zhejiang Normal University, Jinhua 321004, People's Republic of China.}

\keywords{KdV-Calogero-Bogoyavlenskii-Schiff equation; Bell polynomial theory; B{\"a}cklund transformation; Infinite conservation laws; Mixed solutions}

\date{\today}

\begin{document}

\begin{abstract}
Shallow water waves phenomena in nature attract the attention of scholars and play an important role in fields such as tsunamis, tidal waves, solitary waves, and hydraulic engineering. Hereby, for the shallow water waves phenomena in various natural environments, we study the KdV-Calogero-Bogoyavlenskii-Schiff (KdV-CBS) equation. Based on the binary Bell polynomial theory, a new general bilinear B{\"a}cklund transformation, Lax pair and infinite conservation laws of the KdV-CBS equation are derived, and it is proved that it is completely integrable in Lax pair sense. Various types of mixed solutions are constructed by using a combination of Homoclinic test method and symbolic computations. These findings have important significance for the discipline, offering vital insights into the intricate dynamics of the KdV-CBS equation. We hope that our research results could help the researchers understand the nonlinear complex phenomena of the shallow water waves in oceans, rivers and coastal areas. Furthermore, the present work can be directly applied to other nonlinear equations.
\end{abstract}
\maketitle

\section{Introduction}
\label{s:intro}

Nonlinear evolution equations (NLEEs) are widely used in several fields~\cite{1,2,3}, including fluid mechanics, solid physics, plasma bulk wave, biology, quantum mechanics, heat flow phenomena and optical fabrics~\cite{4,5}. The broad range of applications indicated above has piqued the curiosity of researchers, leading to substantial investigation of NLEEs. In order to understand the complex physical phenomena simulated by these NLEEs, it is very important to find their exact solutions~\cite{6,7}. Unlike linear equations, there is no general analytical method and system theory to obtain exact solutions of NLEEs. Over the years, researchers have proposed some effective methods to study the exact solutions of NLEEs~\cite{8,9}, including the inverse scattering transform method~\cite{10,11}, Darboux transformation \cite{12,13,14}, Hirota bilinear method~\cite{15}, B{\"a}cklund transformation method~\cite{16}, long wave limit method~\cite{17}, symmetry analysis method~\cite{18} and many more methods~\cite{19}. Based on these methods, many works have been done to study the analytical solutions of NLEEs~\cite{20,21}.

Lambert~\cite{22} and colleagues proposed a combination of Hirota bilinear method and Bell polynomial method to obtain the bilinear B{\"a}cklund transformations and Lax pair for soliton equations. Applying the binary Bell polynomial method, researchers have not only provided the process of constructing bilinear forms, but also systematically constructed bilinear B{\"a}cklund transformations, Lax pair and infinite conservation laws of NLEEs~\cite{23,24}. The study of conservation laws is very important. Wang~\cite{25} used the zero-curvature equation and Lenard recursive operators to derived the extended Merola-Ragnisco-Tu lattice equation and the infinite conservation laws.

Recently, Wazwaz introduced a new (3+1)-dimensional KdV-Calogero-Bogoyavlenskii-Schiff equation~\cite{26}, which describes the shallow water waves and fluid flow phenomena in various natural environments including oceans, rivers and coastal areas, and has been widely used in physics, mathematics and engineering.
\begin{equation}\label{1.1}
\begin{split}
4u_{t}&+\lambda_{1}(u_{xxy}+4uu_{y}+2u_{x}\partial_{x}^{-1}u_{y})+\lambda_{2}(u_{xxz}+4uu_{z}
+2u_{x}\partial_{x}^{-1}u_{z}) \\
&+\lambda_{3}(u_{xxx}+6uu_{x})+\lambda_{4}u_{x}+\lambda_{5}u_{y}+\lambda_{6}u_{z}=0,
\end{split}
\end{equation}
where $u=u(x,y,z,t)$, with $\lambda_{1}$, $\lambda_{2}$, $\lambda_{3}$, $\lambda_{4}$, $\lambda_{5}$ and $\lambda_{6}$ are real constants. Lump solutions and localized wave solutions of Eq. \eqref{1.1} are constructed by the symbolic computational method~\cite{27}. A set of auto-B{\"a}cklund transformations for Eq. \eqref{1.1} is obtained using truncated Painlev{\'e} expansions, and $N$-soliton solutions for the equation~\cite{28} are derived on a non-zero background. Eq. \eqref{1.1} is simplified into a variety of nonlinear equations with different physical properties by changing its coefficients.
\begin{itemize} \itemsep0.75pt
  \item When $\lambda_{1}=\lambda_{2}=\lambda_{4}=\lambda_{5}=\lambda_{6}=0$ and $\lambda_{3}=4$. Equation \eqref{1.1} has been reduced to the ubiquitous Korteweg-de Vries equation~\cite{29}
\begin{equation}\label{1.2}
\begin{split}
u_{t}+u_{xxx}+6uu_{x}=0.
\end{split}
\end{equation}
  \item When $\lambda_{2}=\lambda_{3}=\lambda_{4}=\lambda_{5}=\lambda_{6}=0$ and $\lambda_{1}=4$. Equation \eqref{1.1} has been reduced to the Calogero-Bogoyavlenskii-Schiff equation~\cite{30}
\begin{equation}\label{1.3}
\begin{split}
u_{t}+u_{xxy}+4uu_{y}+2u_{x}\partial_{x}^{-1}u_{y}=0.
\end{split}
\end{equation}
  \item When $\lambda_{1}=-h_{1}$, $\lambda_{3}=-h_{2}$ and $\lambda_{2}=\lambda_{4}=\lambda_{5}=\lambda_{6}=0$. Equation \eqref{1.1} has been reduced to the generalized (2+1)-dimensional Korteweg-de Vries equation~\cite{31}
\begin{equation}\label{1.4}
\begin{split}
4u_{t}-h_{1}(u_{xxy}+4uu_{y}+2u_{x}\partial_{x}^{-1}u_{y})-h_{2}(u_{xxx}+6uu_{x})=0.
\end{split}
\end{equation}
  \item  Setting $\lambda_{1}=4h_{1}$, $\lambda_{2}=4h_{2}$, $\lambda_{3}=4h_{3}$ and $\lambda_{4}=\lambda_{5}=\lambda_{6}=0$ in Eq. \eqref{1.1} gives the (3+1)-dimensional KdV-like model equation~\cite{32}
\begin{equation}\label{1.5}
\begin{split}
u_{t}+h_{1}(u_{xxy}+4uu_{y}+2u_{x}\partial_{x}^{-1}u_{y})+h_{2}(u_{xxz}+4uu_{z}
+2u_{x}\partial_{x}^{-1}u_{z})+h_{3}(u_{xxx}+6uu_{x})=0.
\end{split}
\end{equation}
\end{itemize}

In recent year, mathematicians have worked hard to develop unique approaches for numerical and analytical solutions to the problems posed by NLEEs. It is worth pointing out that the soliton solutions of other interesting nonlinear integrable flows involving reflection points of coordinates are generated in detail by Riemann-Hilbert problems~\cite{33} or Darboux transformations~\cite{34}. Very recently, various coupled and combined integrable models have been successfully generated from $4\times4$ matrix spectral problems and carefully studied from a bi-Hamiltonian point of view~\cite{35,36}. These interesting integrable models have significant integrability properties and have rich applications in physics, mechanical engineering, and materials science, including plasma physics, condensed state physics and atmospheric oceanography.

Hirota's bilinear method is to study bilinear B{\"a}cklund transformations and some soliton solutions of NLEEs~\cite{37}. The key of Hirota bilinear method is to construct the bilinear form of NLEEs through appropriate variable transformation, and then study its properties. However, for nonlinear evolution equations that cannot be converted into bilinear equations, this method has limitations. Homoclinic test method~\cite{38} can be used to construct mixed solutions of many different functions. By directly assuming different forms to solve with the help of appropriate variable transformation, this method makes up for the deficiency that it cannot be converted into bilinear form. One major benefit of the Homoclinic test method is that it has the potential to generate a greater number of the mixed wave solutions. This enables researchers to get a better understanding of the physical events underpinning the system under consideration and make more precise predictions regarding its behavior.

The paper is organized as follows. In Section~\ref{s:Integrability}, the integrability of Eq. \eqref{1.1} is studied by means of Bell polynomial theory, the B{\"a}cklund transformation, Lax pair and infinite conservation laws of the equation are constructed. In Section~\ref{s:Interactions}, the interactions between lump solutions and multiple functions of Eq. \eqref{1.1} are studied by Homoclinic test method. In Section~\ref{s:Mixed solutions}, the mixed solutions of exponential functions and different functions are presented on our analytic results. The conclusion section gives a few comments and remarks.

\section{Integrability of the KdV-CBS equation}
\label{s:Integrability}

In this section, we propose a new general bilinear B{\"a}cklund transformation Lax pair and infinite conservation laws of new (3+1)-dimensional KdV-CBS equation Eq. \eqref{1.1}, which are obtained from the binary Bell polynomials approach~\cite{39,40}.

\subsection{B{\"a}cklund transformation and Lax pair}

Firstly, we substitute the variable transformation $u=q_{xx}$ into Eq. \eqref{1.1}, the resulting equation can be written as
\begin{equation}\label{2.1}
\begin{split}
4q_{xxt}&+\lambda_{1}(q_{xxxxy}+4q_{xx}q_{xxy}+2q_{xxx}q_{xy})+\lambda_{2}(q_{xxxxz}+4q_{xx}q_{xxz}
+2q_{xxx}q_{xz})  \\
&+\lambda_{3}(q_{xxxxx}+6q_{xx}q_{xxx})+\lambda_{4}q_{xxx}+\lambda_{5}q_{xxy}+\lambda_{6}q_{xxz}=0.
\end{split}
\end{equation}
After integrating once with respect to $x$. Eq. \eqref{2.1} can be written as
\begin{equation}\label{2.2}
\begin{split}
E(q)&=4q_{xt}+\frac{2}{3}\lambda_{1}(q_{xxxy}+3q_{xx}q_{xy})
+\frac{1}{3}\lambda_{1}\partial_{x}^{-1}\partial_{y}(q_{xxxx}+3q_{xx}^{2})
+\frac{1}{3}\lambda_{2}\partial_{x}^{-1}\partial_{z}(q_{xxxx}+3q_{xx}^{2}) \\
&+\frac{2}{3}\lambda_{2}(q_{xxxz}+3q_{xx}q_{xz})
+\lambda_{3}(q_{xxxx}+3q_{xx}^{2})+\lambda_{4}q_{xx}+\lambda_{5}q_{xy}+\lambda_{6}q_{xz}=0.
\end{split}
\end{equation}

Assuming $q=2\ln f$ and $\widetilde{q}=2\ln g$ are two different solutions to Eq. \eqref{2.2}, then we export the following two-field condition:
\begin{equation}\label{2.3}
\begin{split}
E(\widetilde{q})-E(q)&=4(\widetilde{q}-q)_{xt}+\lambda_{1}[(\widetilde{q}-q)_{xxxy}
+(\widetilde{q}-q)_{xx}(\widetilde{q}+q)_{xy}+(\widetilde{q}+q)_{xx}(\widetilde{q}-q)_{xy}] \\
&+\lambda_{1}\partial_{x}^{-1}[(\widetilde{q}-q)_{xx}(\widetilde{q}+q)_{xxy}
+(\widetilde{q}+q)_{xx}(\widetilde{q}-q)_{xxy}]+\lambda_{4}(\widetilde{q}-q)_{xx} \\
&+\lambda_{2}[(\widetilde{q}-q)_{xxxz}+(\widetilde{q}-q)_{xx}(\widetilde{q}+q)_{xz}
+(\widetilde{q}+q)_{xx}(\widetilde{q}-q)_{xz}] \\
&+\lambda_{2}\partial_{x}^{-1}[(\widetilde{q}-q)_{xx}(\widetilde{q}+q)_{xxz}
+(\widetilde{q}+q)_{xx}(\widetilde{q}-q)_{xxz}]+\lambda_{5}(\widetilde{q}-q)_{xy} \\
&+\lambda_{3}[(\widetilde{q}-q)_{xxxx}+3(\widetilde{q}+q)_{xx}(\widetilde{q}-q)_{xx}]
+\lambda_{6}(\widetilde{q}-q)_{xz}=0.
\end{split}
\end{equation}
By taking $\widetilde{q}-q=2v$ and $\widetilde{q}+q=2w$, the expression \eqref{2.3} can be rewritten as
\begin{equation}\label{2.4}
\begin{split}
E(\widetilde{q})-E(q)&=8v_{xt}+2\lambda_{3}(v_{xxxx}+6v_{xx}w_{xx})+2\lambda_{4}v_{xx}+2\lambda_{5}v_{xy}
+2\lambda_{6}v_{xz} \\
&+\lambda_{1}[2v_{xxxy}+4(v_{xx}w_{xy}+w_{xx}v_{xy})+4\partial_{x}^{-1}(v_{xx}w_{xxy}+w_{xx}v_{xxy})] \\
&+\lambda_{2}[2v_{xxxz}+4(v_{xx}w_{xz}+w_{xx}v_{xz})+4\partial_{x}^{-1}(v_{xx}w_{xxz}+w_{xx}v_{xxz})] \\
&=2\partial_{x}[4\mathscr{Y}_{t}(v)+\lambda_{1} \mathscr{Y}_{xxy}(v,w)+\lambda_{2} \mathscr{Y}_{xxz}(v,w)+\lambda_{3} \mathscr{Y}_{3x}(v,w)
+\lambda_{4}\mathscr{Y}_{x}(v)\\
&+\lambda_{5}\mathscr{Y}_{y}(v)+\lambda_{6}\mathscr{Y}_{z}(v)]+\mathscr{R}(v,w)=0,
\end{split}
\end{equation}
where
\begin{equation}\label{2.5}
\begin{split}
\mathscr{R}(v,w)&=4\lambda_{1}[w_{xx}v_{xy}-v_{x}w_{xxy}+\partial_{x}^{-1}(w_{xx}v_{xxy}+v_{xx}w_{xxy})]
-2\lambda_{1} \partial_{x}[v_{y}(v_{x}^{2}+w_{xx})] \\
&+4\lambda_{2}[w_{xx}v_{xz}-v_{x}w_{xxz}+\partial_{x}^{-1}(w_{xx}v_{xxz}+v_{xx}w_{xxz})]-2\lambda_{2} \partial_{x}[v_{z}(v_{x}^{2}+w_{xx})] \\
&-6\lambda_{3} \mathscr{W}[\mathscr{Y}_{x}(v),\mathscr{Y}_{xx}(v,w)],
\end{split}
\end{equation}
with $\mathscr{W}$ is Wronskian.

We set $v_{x}^{2}+w_{xx}=M$, then
\begin{equation}\label{2.6}
\begin{split}
&\mathscr{R}(v,w)=6M \partial_{x}[\lambda_{1} \mathscr{Y}_{y}(v)+\lambda_{2} \mathscr{Y}_{z}(v)+\lambda_{3} \mathscr{Y}_{x}(v)].
\end{split}
\end{equation}
Eq. \eqref{2.4} can be rewritten as a pairs of linear combinations about $\mathscr{Y}$-polynomials
\begin{equation}\label{2.7}
\begin{split}
&\mathscr{Y}_{xx}(v,w)-M=0, \\
&\partial_{x}[4\mathscr{Y}_{t}(v)+\lambda_{1} \mathscr{Y}_{xxy}(v,w)+\lambda_{2} \mathscr{Y}_{xxz}(v,w)+\lambda_{3} \mathscr{Y}_{xxx}(v,w)
+(3M\lambda_{3}+\lambda_{4})\mathscr{Y}_{x}(v) \\
&+(3M\lambda_{1}+\lambda_{5})\mathscr{Y}_{y}(v)+(3M\lambda_{2}+\lambda_{6})\mathscr{Y}_{z}(v)]=0,
\end{split}
\end{equation}
the system \eqref{2.7} leads to the bilinear B{\"a}cklund transformation
\begin{equation}\label{2.8}
\begin{split}
&(D_{x}^{2}-M)f\cdot g=0, \\
&[4D_{t}+\lambda_{3} D_{x}^{3}+\lambda_{1} D_{x}^{2}D_{y}+\lambda_{2} D_{x}^{2}D_{z}+(3M\lambda_{3}+\lambda_{4})D_{x} \\
&+(3M\lambda_{1}+\lambda_{5})D_{y}+(3M\lambda_{2}+\lambda_{6})D_{z} ]f\cdot g=0.
\end{split}
\end{equation}

Based on the Hopf-Cole transformation $v=\ln \psi$, $w=q+\ln \psi$, and linearizing the system \eqref{2.7}, the linear differential equations are obtained as
\begin{equation}\label{2.9}
\begin{split}
&\psi_{xx}+q_{xx}\psi-M\psi=0, \\
&\lambda_{1}(\psi_{xxy}+q_{xx}\psi_{y}+2q_{xy}\psi_{x})
+\lambda_{2}(\psi_{xxz}+q_{xx}\psi_{z}+2q_{xz}\psi_{x})+\lambda_{3}(\psi_{xxx}+3q_{xx}\psi_{x}) \\
&+4\psi_{t}+(3M\lambda_{3}+\lambda_{4})\psi_{x}+(3M\lambda_{1}
+\lambda_{5})\psi_{y}+(3M\lambda_{2}+\lambda_{6})\psi_{z}=0.
\end{split}
\end{equation}
With the compatibility condition $\psi_{xxt}=\psi_{txx}$, we can derive Eq. \eqref{1.1} from Eq. \eqref{2.9}, which means Eq. \eqref{1.1} is completely integrable in Lax pair sense.

\subsection{Infinite conservation laws}
\label{s:Isl}

By introducing a new potential function $\eta=(\widetilde{q}_{x}-q_{x})/2$, then Eq. \eqref{2.7} become
\begin{equation}\label{2.10}
\begin{split}
&\eta_{x}+\eta^{2}+q_{xx}-M=0, \\
&\partial_{t}(4\eta)+\partial_{x}[\lambda_{3}(\eta_{xx}-2\eta^{3})+(6M\lambda_{3}+2\lambda_{1} q_{xy}+2\lambda_{2} q_{xz}+\lambda_{4})\eta] \\
&+\partial_{y}[\lambda_{1}\eta_{xx}+2\lambda_{1}\eta\eta_{x}+(4M\lambda_{1}+\lambda_{5})\eta]
+\partial_{z}[\lambda_{2}\eta_{xx}+2\lambda_{2}\eta\eta_{x}+(4M\lambda_{2}+\lambda_{6})\eta]=0.
\end{split}
\end{equation}
The function $\eta$ and arbitrary parameter $M$ are expanded as the following series
\begin{equation}\label{2.11}
\begin{split}
M=\varepsilon^{2}, \quad \eta=\varepsilon+\sum_{n=1}^{\infty}\mathscr{I}_{n}(q,q_{x},\ldots)\varepsilon^{-n}.
\end{split}
\end{equation}
Inserting series \eqref{2.11} into the first equation of \eqref{2.10} and equating each coefficient for the power of $\varepsilon^{-n}(n=1,2,\ldots)$, we obtain the following recursions:
\begin{equation}\label{2.12}
\begin{split}
&\mathscr{I}_{1}=-\frac{q_{xx}}{2}=-\frac{u}{2}, \quad \mathscr{I}_{2}=-\frac{\mathscr{I}_{1,x}}{2}=\frac{q_{xxx}}{4}=\frac{u_{x}}{4}, \quad
\mathscr{I}_{n+1}=-\frac{1}{2}(\mathscr{I}_{n,x}+\sum_{k=1}^{n}\mathscr{I}_{k}\mathscr{I}_{n-k}), \quad (n=2,3,\ldots).
\end{split}
\end{equation}
Considering the second equation of \eqref{2.10} in the same way, we can get the infinite conservation laws
\begin{equation}\label{2.13}
\begin{split}
\mathscr{G}_{n,t}+\mathscr{M}_{n,x}+\mathscr{N}_{n,y}+\mathscr{F}_{n,z}=0. \quad (n=1,2,3,\ldots).
\end{split}
\end{equation}

The $\mathscr{G}_{n}$ takes the form $\mathscr{G}_{n}=4\mathscr{I}_{n}$, and the first fluxes $\mathscr{M}_{n}$ are given explicitly by the recursion formulas
\begin{equation}\label{2.14}
\begin{split}
\mathscr{M}_{1}&=-\frac{1}{2}(\lambda_{3} u_{xx}+3\lambda_{3}u^{2}+\lambda_{4}u-\lambda_{1}u_{xy}
-\lambda_{2}u_{xz}+2\lambda_{1}u\partial_{x}^{-1}u_{y}+2\lambda_{2}u\partial_{x}^{-1}u_{z}), \\
\mathscr{M}_{2}&=\frac{1}{4}[\lambda_{3}(u_{xxx}+6uu_{x})+\lambda_{4}u_{x}
-\lambda_{1}(u_{xxy}+2uu_{y}-2u_{x}\partial_{x}^{-1}u_{y})
-\lambda_{2}(u_{xxz}+2uu_{z}-2u_{x}\partial_{x}^{-1}u_{z})],  \\
\mathscr{M}_{n}&=\lambda_{3} \mathscr{I}_{n,xx}+6\lambda_{3} \mathscr{I}_{n+2}+\lambda_{4} \mathscr{I}_{n}-2\lambda_{3} \sum_{i+j+k=n}\mathscr{I}_{i}\mathscr{I}_{j}\mathscr{I}_{k}+2\lambda_{1} \mathscr{I}_{n+1,y} \\
&+2\lambda_{2} \mathscr{I}_{n+1,z}+2\lambda_{1} \partial_{x}^{-1}u_{y} \mathscr{I}_{n}+2\lambda_{2} \partial_{x}^{-1}u_{z} \mathscr{I}_{n}, \quad (n=3,4,\ldots).
\end{split}
\end{equation}
The second fluxes $\mathscr{N}_{n}$ are given explicitly by the recursion formulas
\begin{equation}\label{2.15}
\begin{split}
\mathscr{N}_{1}&=-\frac{1}{2}(2\lambda_{1} u_{xx}+\lambda_{1} u^{2}+\lambda_{5}u), \quad
\mathscr{N}_{2}=\frac{1}{2}(\lambda_{1} u_{xxx}+3\lambda_{1} uu_{x}+\frac{\lambda_{5}}{2}u_{x}), \\
\mathscr{N}_{n}&=\lambda_{1} \mathscr{I}_{n,xx}+2\lambda_{1}
\sum_{k=1}^{n}\mathscr{I}_{n-k}\mathscr{I}_{k,x}+4\lambda_{1}\mathscr{I}_{n+2}
+\lambda_{5}\mathscr{I}_{n}, \quad (n=3,4,\ldots).
\end{split}
\end{equation}
And $\mathscr{F}_{n}$ are given by
\begin{equation}\label{2.16}
\begin{split}
\mathscr{F}_{1}&=-\frac{1}{2}(2\lambda_{2} u_{xx}+\lambda_{2} u^{2}+\lambda_{6}u), \quad
\mathscr{F}_{2}=\frac{1}{2}(\lambda_{2} u_{xxx}+3\lambda_{2} uu_{x}+\frac{\lambda_{6}}{2}u_{x}), \\
\mathscr{F}_{n}&=\lambda_{2} \mathscr{I}_{n,xx}+2\lambda_{2} \sum_{k=1}^{n}\mathscr{I}_{n-k}\mathscr{I}_{k,x}
+4\lambda_{2}\mathscr{I}_{n+2}+\lambda_{6}\mathscr{I}_{n}, \quad (n=3,4,\ldots).
\end{split}
\end{equation}
With the recursion formulas $\mathscr{G}_{n}=4\mathscr{I}_{n}$, \eqref{2.14}, \eqref{2.15} and \eqref{2.16} given above, we can construct the infinite conservation laws for Eq. \eqref{1.1}.

\section{Interactions between lump solutions and multiple functions}
\label{s:Interactions}

In this section, we analyze the dynamics of interactions between lump solutions and multiple functions to Eq. \eqref{1.1}.

\subsection{Lump solutions}
\label{s:Lump}

Considering the variable transformation
\begin{equation}\label{3.1}
\begin{split}
u=2(\ln f)_{xx}+u_{0},
\end{split}
\end{equation}
where $u_{0}$ is a non-zero real constant. Eq. \eqref{1.1} has been converted into the following form
\begin{equation}\label{3.2}
\begin{split}
&\lambda_{1}[(\ln f)_{xxxxy}+8(\ln f)_{xx}(\ln f)_{xxy}+4(\ln f)_{xxx}(\ln f)_{xy}]+(4u_{0}\lambda_{1}+\lambda_{5})(\ln f)_{xxy} \\
&+\lambda_{2}[(\ln f)_{xxxxz}+8(\ln f)_{xx}(\ln f)_{xxz}+4(\ln f)_{xxx}(\ln f)_{xz}]+(4u_{0}\lambda_{2}+\lambda_{6})(\ln f)_{xxz} \\
&+4(\ln f)_{xxt}+\lambda_{3}[(\ln f)_{xxxxx}+12(\ln f)_{xx}(\ln f)_{xxx}]+(6u_{0}\lambda_{3}+\lambda_{4})(\ln f)_{xxx}=0.
\end{split}
\end{equation}

In order to search for the test function composed of $N$-quadratic functions for Eq. \eqref{3.2}, we assume that
\begin{equation}\label{3.3}
\begin{split}
f_{A}=m_{0}+\sum_{i=1}^{N}(m_{i}x+n_{i}y+p_{i}z+q_{i}t)^{2},
\end{split}
\end{equation}
where $m_{i}$, $n_{i}$, $p_{i}$, $q_{i}$ are all the real constants, $m_{0}>0$ is required, with $N$ is a positive integer. Substituting the expression \eqref{3.3} into Eq. \eqref{3.2}, the following six cases are obtained.

\textbf{Case~1.1:}
\begin{equation}\label{3.4}
\begin{split}
&m_{i}=\frac{p_{i}(\lambda_{2}\lambda_{5}-\lambda_{1}\lambda_{6})-4\lambda_{1}q_{i}}
{2u_{0}\lambda_{1}\lambda_{3}+\lambda_{1}\lambda_{4}-\lambda_{3}\lambda_{5}}, \quad
n_{i}=\frac{p_{i}(\lambda_{3}\lambda_{6}-\lambda_{2}\lambda_{4}-2u_{0}\lambda_{2}\lambda_{3})
+4\lambda_{3}q_{i}}{2u_{0}\lambda_{1}\lambda_{3}+\lambda_{1}\lambda_{4}-\lambda_{3}\lambda_{5}},
\end{split}
\end{equation}
where $2u_{0}\lambda_{1}\lambda_{3}+\lambda_{1}\lambda_{4}-\lambda_{3}\lambda_{5}\neq0$.

\textbf{Case~1.2:}
\begin{equation}\label{3.5}
\begin{split}
&m_{i}=\frac{n_{i}(\lambda_{1}\lambda_{6}-\lambda_{2}\lambda_{5})-4\lambda_{2}q_{i}}
{2u_{0}\lambda_{2}\lambda_{3}+\lambda_{2}\lambda_{4}-\lambda_{3}\lambda_{6}}, \quad
p_{i}=\frac{n_{i}(\lambda_{3}\lambda_{5}-\lambda_{1}\lambda_{4}-2u_{0}\lambda_{1}\lambda_{3})
+4\lambda_{3}q_{i}}{2u_{0}\lambda_{2}\lambda_{3}+\lambda_{2}\lambda_{4}-\lambda_{3}\lambda_{6}},
\end{split}
\end{equation}
where $2u_{0}\lambda_{2}\lambda_{3}+\lambda_{2}\lambda_{4}-\lambda_{3}\lambda_{6}\neq0$.

\textbf{Case~1.3:}
\begin{equation}\label{3.6}
\begin{split}
m_{i}=-\frac{\lambda_{1}n_{i}+\lambda_{2}p_{i}}{\lambda_{3}}, \quad q_{i}=\frac{n_{i}(2u_{0}\lambda_{1}\lambda_{3}+\lambda_{1}\lambda_{4}-\lambda_{3}\lambda_{5})
+p_{i}(2u_{0}\lambda_{2}\lambda_{3}+\lambda_{2}\lambda_{4}-\lambda_{3}\lambda_{6})}{4\lambda_{3}},
\end{split}
\end{equation}
where $\lambda_{3}\neq0$.

\textbf{Case~1.4:}
\begin{equation}\label{3.7}
\begin{split}
n_{i}=\frac{m_{i}(\lambda_{3}\lambda_{6}-\lambda_{2}\lambda_{4}-2u_{0}\lambda_{2}\lambda_{3})
-4\lambda_{2}q_{i}}{\lambda_{2}\lambda_{5}-\lambda_{1}\lambda_{6}}, \quad
p_{i}=\frac{m_{i}(\lambda_{3}\lambda_{5}-\lambda_{1}\lambda_{4}-2u_{0}\lambda_{1}\lambda_{3})
-4\lambda_{1}q_{i}}{\lambda_{1}\lambda_{6}-\lambda_{2}\lambda_{5}},
\end{split}
\end{equation}
where $\lambda_{2}\lambda_{5}-\lambda_{1}\lambda_{6}\neq0$.

\textbf{Case~1.5:}
\begin{equation}\label{3.8}
\begin{split}
n_{i}=-\frac{\lambda_{3}m_{i}+\lambda_{2}p_{i}}{\lambda_{1}}, \quad
q_{i}=\frac{m_{i}(\lambda_{3}\lambda_{5}-\lambda_{1}\lambda_{4}-2u_{0}\lambda_{1}\lambda_{3})
+p_{i}(\lambda_{2}\lambda_{5}-\lambda_{1}\lambda_{6})}{4\lambda_{1}},
\end{split}
\end{equation}
where $\lambda_{1}\neq0$.

\textbf{Case~1.6:}
\begin{equation}\label{3.9}
\begin{split}
p_{i}=-\frac{\lambda_{1}n_{i}+\lambda_{3}m_{i}}{\lambda_{2}}, \quad
q_{i}=\frac{m_{i}(\lambda_{3}\lambda_{6}-\lambda_{2}\lambda_{4}-2u_{0}\lambda_{2}\lambda_{3})
+n_{i}(\lambda_{1}\lambda_{6}-\lambda_{2}\lambda_{5})}{4\lambda_{2}},
\end{split}
\end{equation}
where $\lambda_{2}\neq0$.

Substituting relational formula $f_{A}$ \eqref{3.3} and the expression \eqref{3.8} into transformation \eqref{3.1}, corresponding lump solutions for Eq. \eqref{1.1} appear as
\begin{equation}\label{3.10}
\begin{split}
u_{A}&=2(\ln f_{A})_{xx}+u_{0}, \quad f_{A}=m_{0}+\sum_{i=1}^{N}(m_{i}x+n_{i}y+p_{i}z+q_{i}t)^{2}, \\
n_{i}&=-\frac{\lambda_{3}m_{i}+\lambda_{2}p_{i}}{\lambda_{1}}, \quad
q_{i}=\frac{m_{i}(\lambda_{3}\lambda_{5}-\lambda_{1}\lambda_{4}-2u_{0}\lambda_{1}\lambda_{3})
+p_{i}(\lambda_{2}\lambda_{5}-\lambda_{1}\lambda_{6})}{4\lambda_{1}}.
\end{split}
\end{equation}

\begin{figure}[htb]
\centering
\includegraphics[width=0.325\textwidth]{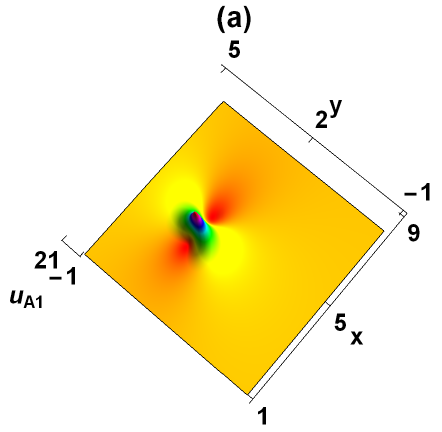}
\includegraphics[width=0.325\textwidth]{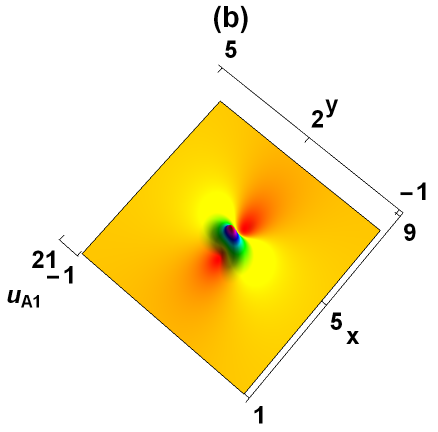}
\includegraphics[width=0.325\textwidth]{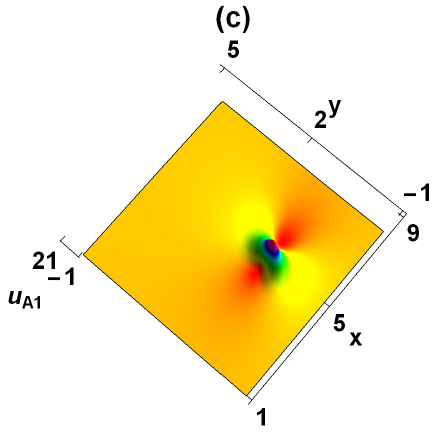}
\caption{\small Profile of $u_{A1}$ in the lump solutions \eqref{3.10} with parameters: $u_{0}=\lambda_{1}=\lambda_{3}=\lambda_{5}=m_{0}=1$, $\lambda_{2}=\lambda_{4}=\lambda_{6}=m_{1}=2$, $m_{2}=0.6$, $m_{3}=-0.8$, $p_{1}=-1.5$, $p_{2}=1.2$, $p_{3}=-0.5$. (a) $z=6$, $t=2$, (b) $z=t=4$, and (c) $z=2$, $t=6$.
}\label{fig:1}
\end{figure}

When $N=3$ is selected among the lump solutions \eqref{3.10}, we noticed that $u_{A1}$ has an upward peak and two downward valleys in Fig.~\ref{fig:1}. This form of the lump structure is called bright lump structure, it can be seen from Fig.~\ref{fig:1} that the energy distribution of bright solitons is obvious. The amplitude and shape did not change during this process.

\subsection{Multiple mixed function solutions}

The mixed solutions of different functions can describe the nonlinear phenomena in nature, so we study the solutions formed by the combination of many different functions.
\begin{equation}\label{3.11}
\begin{split}
f_{B}=\sum_{j=1}^{M}k_{j}\Delta_{j}^{\rho_{j}}(a_{j}x+b_{j}y+c_{j}z+d_{j}t),
\end{split}
\end{equation}
where $k_{j}$, $a_{j}$, $b_{j}$, $c_{j}$, $d_{j}$ are all the real constants, $\rho_{j}$ is an integer. $\Delta_{j}$ is any elementary function, including the trigonometric function, hyperbolic function, Jacobian elliptic function and inverse trigonometric function. Substituting expressions \eqref{3.11} into Eq. \eqref{3.2}, we get the following six cases:

\textbf{Case~2.1:}
\begin{equation}\label{3.12}
\begin{split}
b_{j}&=\frac{a_{j}(2u_{0}\lambda_{2}\lambda_{3}+\lambda_{2}\lambda_{4}-\lambda_{3}\lambda_{6})
+4\lambda_{2}d_{j}}{\lambda_{1}\lambda_{6}-\lambda_{2}\lambda_{5}}, \quad
c_{j}=\frac{a_{j}(2u_{0}\lambda_{1}\lambda_{3}+\lambda_{1}\lambda_{4}-\lambda_{3}\lambda_{5})
+4\lambda_{1}d_{j}}{\lambda_{2}\lambda_{5}-\lambda_{1}\lambda_{6}},
\end{split}
\end{equation}
where $\lambda_{1}\lambda_{6}-\lambda_{2}\lambda_{5}\neq0$.

\textbf{Case~2.2:}
\begin{equation}\label{3.13}
\begin{split}
b_{j}=-\frac{\lambda_{3}a_{j}+\lambda_{2}c_{j}}{\lambda_{1}}, \quad
d_{j}=\frac{a_{j}(\lambda_{3}\lambda_{5}-\lambda_{1}\lambda_{4}-2u_{0}\lambda_{1}\lambda_{3})
+c_{j}(\lambda_{2}\lambda_{5}-\lambda_{1}\lambda_{6})}{4\lambda_{1}},
\end{split}
\end{equation}
where $\lambda_{1}\neq0$.

\textbf{Case~2.3:}
\begin{equation}\label{3.14}
\begin{split}
c_{j}=-\frac{\lambda_{1}b_{j}+\lambda_{3}a_{j}}{\lambda_{2}}, \quad
d_{j}=\frac{a_{j}(\lambda_{3}\lambda_{6}-\lambda_{2}\lambda_{4}-2u_{0}\lambda_{2}\lambda_{3})
+b_{j}(\lambda_{1}\lambda_{6}-\lambda_{2}\lambda_{5})}{4\lambda_{2}},
\end{split}
\end{equation}
where $\lambda_{2}\neq0$.

\textbf{Case~2.4:}
\begin{equation}\label{3.15}
\begin{split}
a_{j}=-\frac{\lambda_{1}b_{j}+\lambda_{2}c_{j}}{\lambda_{3}}, \quad
d_{j}=\frac{b_{j}(2u_{0}\lambda_{1}\lambda_{3}+\lambda_{1}\lambda_{4}-\lambda_{3}\lambda_{5})
+c_{j}(2u_{0}\lambda_{2}\lambda_{3}+\lambda_{2}\lambda_{4}-\lambda_{3}\lambda_{6})}{4\lambda_{3}},
\end{split}
\end{equation}
where $\lambda_{3}\neq0$.

\textbf{Case~2.5:}
\begin{equation}\label{3.16}
\begin{split}
a_{j}=\frac{c_{j}(\lambda_{2}\lambda_{5}-\lambda_{1}\lambda_{6})-4\lambda_{1}d_{j}}
{2u_{0}\lambda_{1}\lambda_{3}+\lambda_{1}\lambda_{4}-\lambda_{3}\lambda_{5}}, \quad
b_{j}=\frac{c_{j}(\lambda_{3}\lambda_{6}-\lambda_{2}\lambda_{4}-2u_{0}\lambda_{2}\lambda_{3})
+4\lambda_{3}d_{j}}{2u_{0}\lambda_{1}\lambda_{3}+\lambda_{1}\lambda_{4}-\lambda_{3}\lambda_{5}},
\end{split}
\end{equation}
where $2u_{0}\lambda_{1}\lambda_{3}+\lambda_{1}\lambda_{4}-\lambda_{3}\lambda_{5}\neq0$.

\textbf{Case~2.6:}
\begin{equation}\label{3.17}
\begin{split}
a_{j}=\frac{b_{j}(\lambda_{1}\lambda_{6}-\lambda_{2}\lambda_{5})-4\lambda_{2}d_{j}}
{2u_{0}\lambda_{2}\lambda_{3}+\lambda_{2}\lambda_{4}-\lambda_{3}\lambda_{6}}, \quad
c_{j}=\frac{b_{j}(\lambda_{3}\lambda_{5}-\lambda_{1}\lambda_{4}-2u_{0}\lambda_{1}\lambda_{3})
+4\lambda_{3}d_{j}}{2u_{0}\lambda_{2}\lambda_{3}+\lambda_{2}\lambda_{4}-\lambda_{3}\lambda_{6}},
\end{split}
\end{equation}
where $2u_{0}\lambda_{2}\lambda_{3}+\lambda_{2}\lambda_{4}-\lambda_{3}\lambda_{6}\neq0$.

Combining expression \eqref{3.17} and \eqref{3.11}, we get the multiple mixed function solutions of Eq. \eqref{1.1}.
\begin{equation}\label{3.18}
\begin{split}
u_{B}&=2(\ln f_{B})_{xx}+u_{0}, \quad f_{B}=\sum_{j=1}^{M}k_{j}\Delta_{j}^{\rho_{j}}(a_{j}x+b_{j}y+c_{j}z+d_{j}t), \\
a_{j}&=\frac{b_{j}(\lambda_{1}\lambda_{6}-\lambda_{2}\lambda_{5})-4\lambda_{2}d_{j}}
{2u_{0}\lambda_{2}\lambda_{3}+\lambda_{2}\lambda_{4}-\lambda_{3}\lambda_{6}}, \quad
c_{j}=\frac{b_{j}(\lambda_{3}\lambda_{5}-\lambda_{1}\lambda_{4}-2u_{0}\lambda_{1}\lambda_{3})
+4\lambda_{3}d_{j}}{2u_{0}\lambda_{2}\lambda_{3}+\lambda_{2}\lambda_{4}-\lambda_{3}\lambda_{6}}.
\end{split}
\end{equation}

\begin{figure}[htb]
\centering
\includegraphics[width=0.325\textwidth]{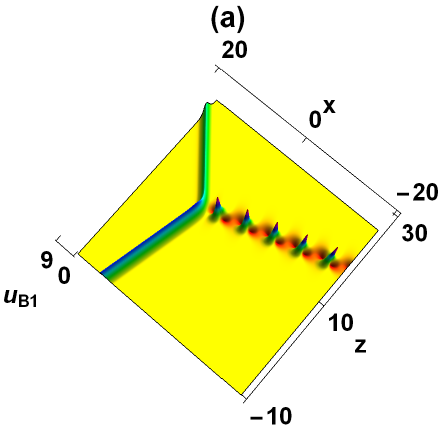}
\includegraphics[width=0.325\textwidth]{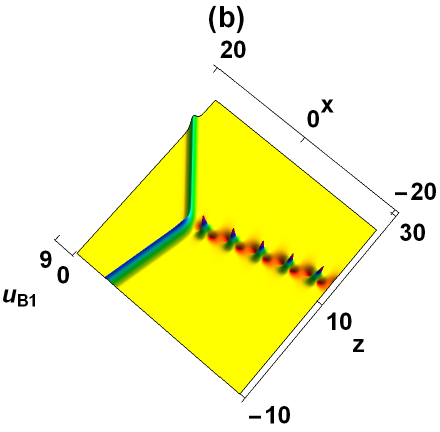}
\includegraphics[width=0.325\textwidth]{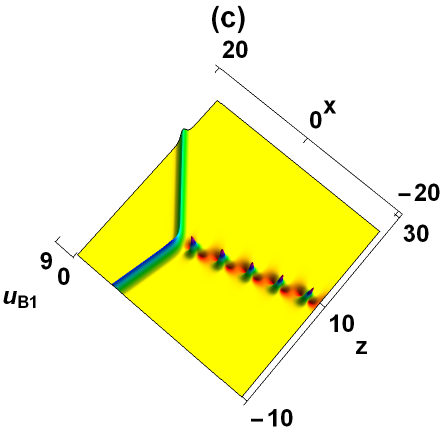}
\caption{\small Interactions between one curved bell shaped wave and one breather soliton via the multiple mixed function solutions \eqref{3.18} with $u_{0}=\lambda_{1}=\lambda_{3}=\lambda_{5}=k_{3}=1$, $\lambda_{2}=\lambda_{4}=\lambda_{6}=k_{1}=2$, $k_{2}=1.5$, $b_{1}=3$, $b_{2}=d_{1}=0.3$, $d_{2}=0.8$, $b_{3}=-1.5$, $d_{3}=-2$. (a) $y=6$, $t=2$, (b) $y=t=4$, and (c) $y=2$, $t=6$.
}\label{fig:2}
\end{figure}

Taking $M=3$, $\rho_{1}=\rho_{2}=\rho_{3}=1$, $\Delta_{1}=\cosh$, $\Delta_{2}=\cos$ and $\Delta_{3}=\exp$ in the multiple mixed function solutions \eqref{3.18}, we derive the solutions which can describe the interactions between one curved bell shaped wave and one breather soliton. In the process of collision with breather soliton, the bell shaped wave deforms and the bending angle of the bell shaped wave increases. It can be seen from Fig.~\ref{fig:2} that the curved bell shaped wave and the breather soliton propagate steadily in a straight line along the negative direction of the $z$-axis, and the amplitude and shape of the curved bell shaped wave and the breather soliton remain unchanged during the movement.

Taking $\rho_{j}=1$ and $\Delta_{j}=\cosh$ in expressions \eqref{3.11} and \eqref{3.15}, we discuss the multiple-cosh soliton solutions formed by a combination of $M$-cosh expressions:
\begin{equation}\label{3.19}
\begin{split}
u_{B}&=2(\ln f_{B})_{xx}+u_{0}, \quad f_{B}=\sum_{j=1}^{M}k_{j}\cosh(a_{j}x+b_{j}y+c_{j}z+d_{j}t),  \\
a_{j}&=-\frac{\lambda_{1}b_{j}+\lambda_{2}c_{j}}{\lambda_{3}}, \quad
d_{j}=\frac{b_{j}(2u_{0}\lambda_{1}\lambda_{3}+\lambda_{1}\lambda_{4}-\lambda_{3}\lambda_{5})
+c_{j}(2u_{0}\lambda_{2}\lambda_{3}+\lambda_{2}\lambda_{4}-\lambda_{3}\lambda_{6})}{4\lambda_{3}}.
\end{split}
\end{equation}

\begin{figure}[htb]
\centering
\includegraphics[width=0.325\textwidth]{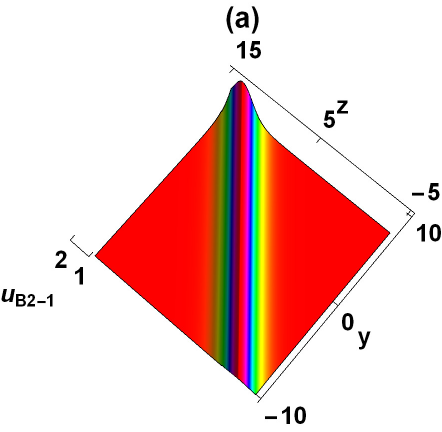}
\includegraphics[width=0.325\textwidth]{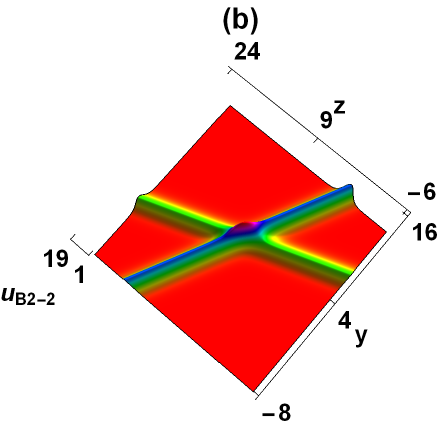}
\includegraphics[width=0.325\textwidth]{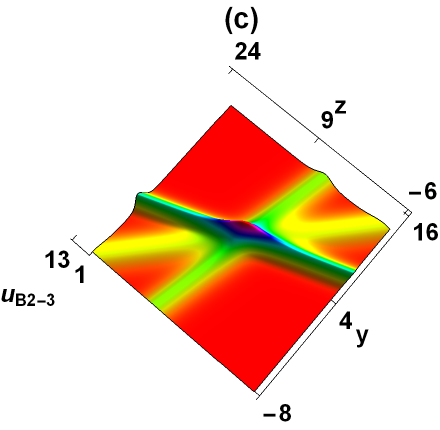}
\caption{\small One bell shaped wave solution (a) with parameters: $u_{0}=\lambda_{1}=\lambda_{3}=\lambda_{5}=1$, $\lambda_{2}=\lambda_{4}=\lambda_{6}=2$, $k_{1}=1$, $b_{1}=0.5$, $c_{1}=-0.6$, $x=6$, $t=2$. Two bell shaped waves solution (b) with parameters: $u_{0}=\lambda_{1}=\lambda_{3}=\lambda_{5}=1$, $\lambda_{2}=\lambda_{4}=\lambda_{6}=2$, $k_{1}=1$, $b_{1}=0.5$, $c_{1}=-0.6$, $k_{2}=1.2$, $b_{2}=-1.6$, $c_{2}=-1.2$, $x=6$, $t=2$. Three bell shaped waves solution (c) with parameters: $u_{0}=\lambda_{1}=\lambda_{3}=\lambda_{5}=1$, $\lambda_{2}=\lambda_{4}=\lambda_{6}=2$, $k_{1}=1$, $b_{1}=0.5$, $c_{1}=-0.6$, $k_{2}=1.2$, $b_{2}=-1.6$, $c_{2}=-1.2$, $k_{3}=1.5$, $b_{3}=-1.3$, $c_{3}=-0.1$, $x=6$, $t=2$.
}\label{fig:3}
\end{figure}

In Fig.~\ref{fig:3} (a), the bell shaped wave determined by the multiple-cosh soliton solutions \eqref{3.19} at $M=1$ is shown, which presents a smooth surface. Fig.~\ref{fig:3} (b) shows the two bell shaped waves determined by the multiple-cosh soliton solutions \eqref{3.19} at $M=2$, and the unimodal soliton formed by the collision of the two bell shaped waves. Fig.~\ref{fig:3} (c) simulates three bell shaped waves determined by the multiple-cosh soliton solutions \eqref{3.19} at $M=3$, and the collision produces the unimodal soliton. The peak value produced by three bell waves is significantly smaller than that produced by two bell waves. Through comparison, it can be found that the peak value of the unimodal soliton produced by the collision of multiple bell waves decreases gradually.

\subsection{Hybrid solutions composed of lump solitons and multiple functions}

To derive the hybrid solutions composed of lump solitons and multiple functions for Eq. \eqref{1.1}, we set
\begin{equation}\label{3.20}
\begin{split}
f_{C}&=m_{0}+\sum_{i=1}^{N}(m_{i}x+n_{i}y+p_{i}z+q_{i}t)^{2}
+\sum_{j=1}^{M}k_{j}\Delta_{j}^{\rho_{j}}(a_{j}x+b_{j}y+c_{j}z+d_{j}t),
\end{split}
\end{equation}
where $m_{i}$, $n_{i}$, $p_{i}$, $q_{i}$, $k_{j}$, $a_{j}$, $b_{j}$, $c_{j}$, $d_{j}$ are all the real constants, $m_{0}>0$ is required, with $\rho_{j}$ is an integer, $\Delta_{j}$ is any elementary function.

\begin{theorem}
Hybrid solutions composed of lump solitons and multiple functions $f_{C}$ \eqref{3.20} are composed of $N$ quadratic functions \eqref{3.3} and multiple mixed function solutions \eqref{3.11}. If the hybrid test function $f_{C}$ \eqref{3.20} is the solution of the Eq. \eqref{3.2}, then any term in conditions \eqref{3.4}, \eqref{3.5}, \eqref{3.6}, \eqref{3.7}, \eqref{3.8}, \eqref{3.9} and any term in conditions \eqref{3.12}, \eqref{3.13}, \eqref{3.14}, \eqref{3.15}, \eqref{3.16}, \eqref{3.17} must be satisfied. There are $36$ cases holds, we only choose the following two cases.
\end{theorem}

\textbf{Case~3.1:}
\begin{equation}\label{3.21}
\begin{split}
p_{i}&=-\frac{\lambda_{1}n_{i}+\lambda_{3}m_{i}}{\lambda_{2}}, \quad
q_{i}=\frac{m_{i}(\lambda_{3}\lambda_{6}-\lambda_{2}\lambda_{4}-2u_{0}\lambda_{2}\lambda_{3})
+n_{i}(\lambda_{1}\lambda_{6}-\lambda_{2}\lambda_{5})}{4\lambda_{2}}, \\
b_{j}&=-\frac{\lambda_{3}a_{j}+\lambda_{2}c_{j}}{\lambda_{1}}, \quad
d_{j}=\frac{a_{j}(\lambda_{3}\lambda_{5}-\lambda_{1}\lambda_{4}-2u_{0}\lambda_{1}\lambda_{3})
+c_{j}(\lambda_{2}\lambda_{5}-\lambda_{1}\lambda_{6})}{4\lambda_{1}},
\end{split}
\end{equation}
where $\lambda_{1}\lambda_{2}\neq0$.

\textbf{Case~3.2:}
\begin{equation}\label{3.22}
\begin{split}
m_{i}&=\frac{p_{i}(\lambda_{2}\lambda_{5}-\lambda_{1}\lambda_{6})-4\lambda_{1}q_{i}}
{2u_{0}\lambda_{1}\lambda_{3}+\lambda_{1}\lambda_{4}-\lambda_{3}\lambda_{5}}, \quad
n_{i}=\frac{p_{i}(\lambda_{3}\lambda_{6}-\lambda_{2}\lambda_{4}-2u_{0}\lambda_{2}\lambda_{3})
+4\lambda_{3}q_{i}}{2u_{0}\lambda_{1}\lambda_{3}+\lambda_{1}\lambda_{4}-\lambda_{3}\lambda_{5}}, \\
b_{j}&=\frac{a_{j}(2u_{0}\lambda_{2}\lambda_{3}+\lambda_{2}\lambda_{4}-\lambda_{3}\lambda_{6})
+4\lambda_{2}d_{j}}{\lambda_{1}\lambda_{6}-\lambda_{2}\lambda_{5}}, \quad
c_{j}=\frac{a_{j}(2u_{0}\lambda_{1}\lambda_{3}+\lambda_{1}\lambda_{4}-\lambda_{3}\lambda_{5})
+4\lambda_{1}d_{j}}{\lambda_{2}\lambda_{5}-\lambda_{1}\lambda_{6}},
\end{split}
\end{equation}
where $\lambda_{1}\lambda_{6}-\lambda_{2}\lambda_{5}\neq0$ and $2u_{0}\lambda_{1}\lambda_{3}+\lambda_{1}\lambda_{4}-\lambda_{3}\lambda_{5}\neq0$.

Combining expression \eqref{3.20} and \eqref{3.21}, we get the hybrid solutions composed of lump solitons and multiple functions of Eq. \eqref{1.1}.
\begin{equation}\label{3.23}
\begin{split}
u_{C}&=2(\ln f_{C})_{xx}+u_{0}, \quad f_{C}=m_{0}+\sum_{i=1}^{N}(m_{i}x+n_{i}y+p_{i}z+q_{i}t)^{2}
+\sum_{j=1}^{M}k_{j}\Delta_{j}^{\rho_{j}}(\theta_{j}), \\
p_{i}&=-\frac{\lambda_{1}n_{i}+\lambda_{3}m_{i}}{\lambda_{2}}, \quad
q_{i}=\frac{m_{i}(\lambda_{3}\lambda_{6}-\lambda_{2}\lambda_{4}-2u_{0}\lambda_{2}\lambda_{3})
+n_{i}(\lambda_{1}\lambda_{6}-\lambda_{2}\lambda_{5})}{4\lambda_{2}},  \\
\theta_{j}&=a_{j}x-\frac{\lambda_{3}a_{j}+\lambda_{2}c_{j}}{\lambda_{1}}y+c_{j}z
+\frac{a_{j}(\lambda_{3}\lambda_{5}-\lambda_{1}\lambda_{4}-2u_{0}\lambda_{1}\lambda_{3})
+c_{j}(\lambda_{2}\lambda_{5}-\lambda_{1}\lambda_{6})}{4\lambda_{1}}t.
\end{split}
\end{equation}

\begin{figure}[htb]
\centering
\includegraphics[width=0.325\textwidth]{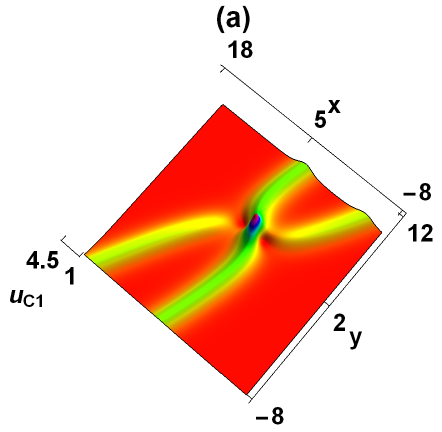}
\includegraphics[width=0.325\textwidth]{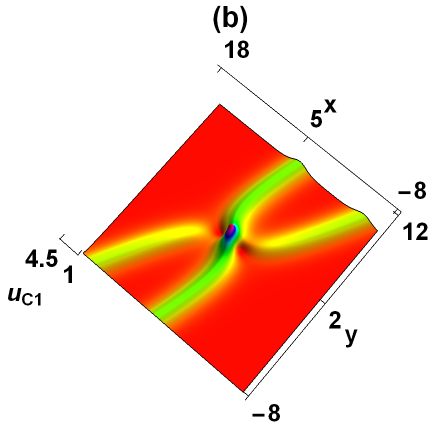}
\includegraphics[width=0.325\textwidth]{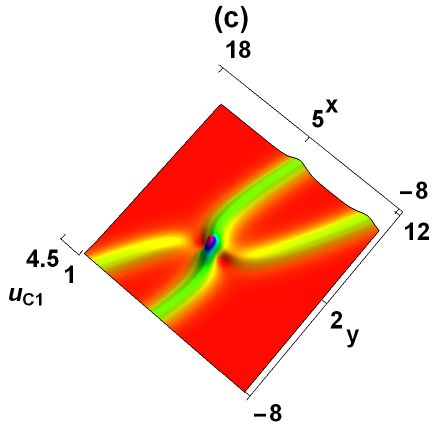}
\caption{\small Interactions between two bell shaped waves and the lump soliton via the hybrid solutions composed of lump solitons and multiple functions \eqref{3.23} with $u_{0}=\lambda_{1}=\lambda_{3}=\lambda_{5}=k_{2}=c_{1}=1$, $\lambda_{2}=\lambda_{4}=\lambda_{6}=k_{1}=n_{2}=2$, $m_{0}=2.5$, $m_{1}=0.6$, $n_{1}=2.1$, $m_{2}=-1.4$, $a_{1}=-1.3$, $a_{2}=1.3$, $c_{2}=-1$. (a) $z=8$, $t=0$, (b) $z=t=4$, and (c) $z=0$, $t=8$.
}\label{fig:4}
\end{figure}

Taking $N=M=2$, $\rho_{1}=\rho_{2}=1$ and $\Delta_{1}=\Delta_{2}=\exp$ in the hybrid solutions composed of lump solitons and multiple functions \eqref{3.23}, we derive the solutions which can describe the interactions between two bell shaped waves and the lump soliton. It can be seen from Fig.~\ref{fig:4} that the lump soliton is distributed at the intersection of two bell shaped waves. Under the interaction of lump soliton, two bell shaped waves deform at the intersection.

Substituting relational formula $f_{C}$ \eqref{3.20} and the expression \eqref{3.22} into transformation \eqref{3.1}, another hybrid solutions composed of lump solitons and multiple functions for Eq. \eqref{1.1} appear as
\begin{equation}\label{3.24}
\begin{split}
u_{C}&=2(\ln f_{C})_{xx}+u_{0}, \quad
f_{C}=m_{0}+\sum_{i=1}^{N}(m_{i}x+n_{i}y+p_{i}z+q_{i}t)^{2}
+\sum_{j=1}^{M}k_{j}\Delta_{j}^{\rho_{j}}(a_{j}x+b_{j}y+c_{j}z+d_{j}t), \\
m_{i}&=\frac{p_{i}(\lambda_{2}\lambda_{5}-\lambda_{1}\lambda_{6})-4\lambda_{1}q_{i}}
{2u_{0}\lambda_{1}\lambda_{3}+\lambda_{1}\lambda_{4}-\lambda_{3}\lambda_{5}}, \quad
n_{i}=\frac{p_{i}(\lambda_{3}\lambda_{6}-\lambda_{2}\lambda_{4}-2u_{0}\lambda_{2}\lambda_{3})
+4\lambda_{3}q_{i}}{2u_{0}\lambda_{1}\lambda_{3}+\lambda_{1}\lambda_{4}-\lambda_{3}\lambda_{5}}, \\
c_{j}&=\frac{a_{j}(2u_{0}\lambda_{1}\lambda_{3}+\lambda_{1}\lambda_{4}-\lambda_{3}\lambda_{5})
+4\lambda_{1}d_{j}}{\lambda_{2}\lambda_{5}-\lambda_{1}\lambda_{6}},  \quad b_{j}=\frac{a_{j}(2u_{0}\lambda_{2}\lambda_{3}+\lambda_{2}\lambda_{4}-\lambda_{3}\lambda_{6})
+4\lambda_{2}d_{j}}{\lambda_{1}\lambda_{6}-\lambda_{2}\lambda_{5}}.
\end{split}
\end{equation}

\begin{figure}[htb]
\centering
\includegraphics[width=0.325\textwidth]{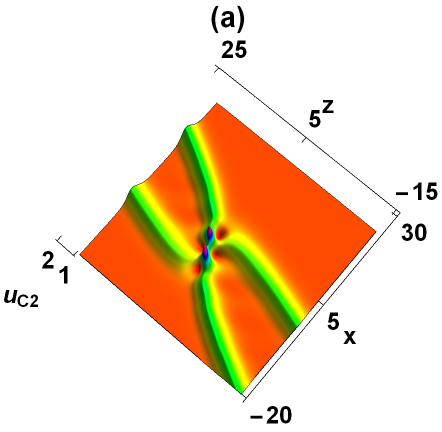}
\includegraphics[width=0.325\textwidth]{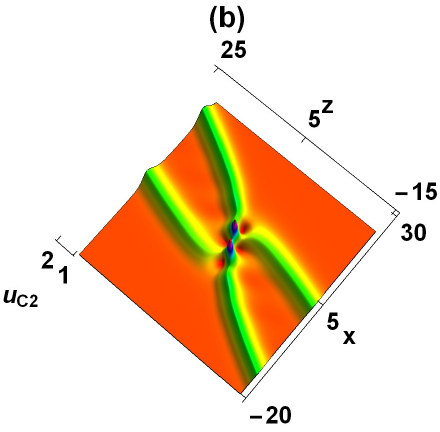}
\includegraphics[width=0.325\textwidth]{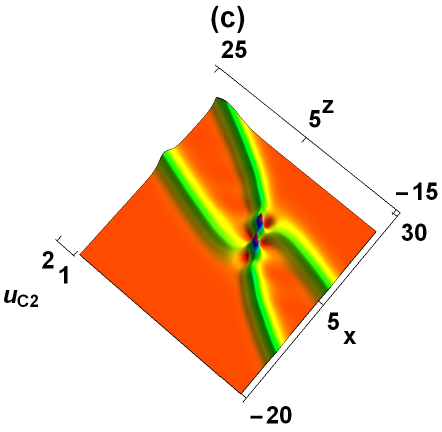}
\caption{\small Interactions between two bell shaped waves and two lump solitons via the hybrid solutions composed of lump solitons and multiple functions \eqref{3.24} with $u_{0}=\lambda_{3}=\lambda_{5}=k_{1}=1$, $\lambda_{1}=3$, $\lambda_{2}=\lambda_{4}=\lambda_{6}=2$, $m_{0}=1.5$, $p_{1}=1.1$, $q_{1}=0.3$, $p_{2}=-1.8$, $q_{2}=0.5$, $k_{2}=1.3$, $a_{1}=0.8$, $d_{1}=-0.6$, $a_{2}=-0.8$, $d_{2}=0.9$. (a) $y=12$, $t=-4$, (b) $y=t=4$, and (c) $y=-4$, $t=12$.
}\label{fig:5}
\end{figure}

Taking $N=M=2$, $\rho_{1}=\rho_{2}=1$, $\Delta_{1}=\cosh$ and $\Delta_{2}=\cos$ in the hybrid solutions composed of lump solitons and multiple functions \eqref{3.24}, we derive the solutions which can describe the interactions between two bell shaped waves and two lump solitons. In Fig.~\ref{fig:5}, it is observed that two lump solitons have two upward peaks and four downward valleys, and distributed at the intersection of two bell shaped waves.

\section{Mixed solutions of exponential functions and different functions}
\label{s:Mixed solutions}

We will study the interaction between exponential functions and different functions, as well as the interaction between the products of different functions.

\subsection{Interaction solutions of exponential functions and multiple mixed functions}

In order to obtain the interaction solutions of exponential functions and multiple mixed functions can be given as follows:
\begin{equation}\label{4.1}
\begin{split}
f_{D}&=r_{0}\exp(\xi)+r_{5}\exp(-\xi)+\sum_{i=1}^{N}k_{i}\Lambda_{i}^{\rho_{i}}(\varphi_{i}), \\
\xi&=r_{1}x+r_{2}y+r_{3}z+r_{4}t, \quad \varphi_{i}=\alpha_{i}x+\beta_{i}y+\gamma_{i}z+\omega_{i}t,
\end{split}
\end{equation}
where $r_{0}$, $r_{1}$, $r_{2}$, $r_{3}$, $r_{4}$, $r_{5}$, $k_{i}$, $\alpha_{i}$, $\beta_{i}$, $\gamma_{i}$, $\omega_{i}$ are all the real constants to be determined, $\rho_{i}$ is an integer, $\Lambda_{i}$ is any elementary function. Substituting expressions \eqref{4.1} into Eq. \eqref{3.2}, we give the following results:

\textbf{Case~4.1:}
\begin{equation}\label{4.2}
\begin{split}
r_{1}&=\frac{r_{3}(\lambda_{2}\lambda_{5}-\lambda_{1}\lambda_{6})-4\lambda_{1}r_{4}}
{2u_{0}\lambda_{1}\lambda_{3}+\lambda_{1}\lambda_{4}-\lambda_{3}\lambda_{5}}, \quad
r_{2}=\frac{r_{3}(\lambda_{3}\lambda_{6}-\lambda_{2}\lambda_{4}-2u_{0}\lambda_{2}\lambda_{3})
+4\lambda_{3}r_{4}}{2u_{0}\lambda_{1}\lambda_{3}+\lambda_{1}\lambda_{4}-\lambda_{3}\lambda_{5}}, \\
\gamma_{i}&=-\frac{\lambda_{1}\beta_{i}+\lambda_{3}\alpha_{i}}{\lambda_{2}}, \quad
\omega_{i}=\frac{\alpha_{i}(\lambda_{3}\lambda_{6}-\lambda_{2}\lambda_{4}-2u_{0}\lambda_{2}\lambda_{3})
+\beta_{i}(\lambda_{1}\lambda_{6}-\lambda_{2}\lambda_{5})}{4\lambda_{2}}.
\end{split}
\end{equation}

\textbf{Case~4.2:}
\begin{equation}\label{4.3}
\begin{split}
r_{1}&=\frac{r_{2}(\lambda_{1}\lambda_{6}-\lambda_{2}\lambda_{5})-4\lambda_{2}r_{4}}
{2u_{0}\lambda_{2}\lambda_{3}+\lambda_{2}\lambda_{4}-\lambda_{3}\lambda_{6}}, \quad
r_{3}=\frac{r_{2}(\lambda_{3}\lambda_{5}-\lambda_{1}\lambda_{4}-2u_{0}\lambda_{1}\lambda_{3})
+4\lambda_{3}r_{4}}{2u_{0}\lambda_{2}\lambda_{3}+\lambda_{2}\lambda_{4}-\lambda_{3}\lambda_{6}}, \\
\beta_{i}&=-\frac{\lambda_{3}\alpha_{i}+\lambda_{2}\gamma_{i}}{\lambda_{1}}, \quad
\omega_{i}=\frac{\alpha_{i}(\lambda_{3}\lambda_{5}-\lambda_{1}\lambda_{4}-2u_{0}\lambda_{1}\lambda_{3})
+\gamma_{i}(\lambda_{2}\lambda_{5}-\lambda_{1}\lambda_{6})}{4\lambda_{1}}.
\end{split}
\end{equation}

\textbf{Case~4.3:}
\begin{equation}\label{4.4}
\begin{split}
r_{1}&=-\frac{\lambda_{1}r_{2}+\lambda_{2}r_{3}}{\lambda_{3}}, \quad
r_{4}=\frac{r_{2}(2u_{0}\lambda_{1}\lambda_{3}+\lambda_{1}\lambda_{4}-\lambda_{3}\lambda_{5})
+r_{3}(2u_{0}\lambda_{2}\lambda_{3}+\lambda_{2}\lambda_{4}-\lambda_{3}\lambda_{6})}{4\lambda_{3}}, \\
\beta_{i}&=\frac{\alpha_{i}(2u_{0}\lambda_{2}\lambda_{3}+\lambda_{2}\lambda_{4}-\lambda_{3}\lambda_{6})
+4\lambda_{2}\omega_{i}}{\lambda_{1}\lambda_{6}-\lambda_{2}\lambda_{5}}, \quad
\gamma_{i}=\frac{\alpha_{i}(2u_{0}\lambda_{1}\lambda_{3}+\lambda_{1}\lambda_{4}-\lambda_{3}\lambda_{5})
+4\lambda_{1}\omega_{i}}{\lambda_{2}\lambda_{5}-\lambda_{1}\lambda_{6}}.
\end{split}
\end{equation}

\textbf{Case~4.4:}
\begin{equation}\label{4.5}
\begin{split}
r_{2}&=\frac{r_{1}(2u_{0}\lambda_{2}\lambda_{3}+\lambda_{2}\lambda_{4}-\lambda_{3}\lambda_{6})
+4\lambda_{2}r_{4}}{\lambda_{1}\lambda_{6}-\lambda_{2}\lambda_{5}}, \quad
r_{3}=\frac{r_{1}(2u_{0}\lambda_{1}\lambda_{3}+\lambda_{1}\lambda_{4}-\lambda_{3}\lambda_{5})
+4\lambda_{1}r_{4}}{\lambda_{2}\lambda_{5}-\lambda_{1}\lambda_{6}}, \\
\alpha_{i}&=-\frac{\lambda_{1}\beta_{i}+\lambda_{2}\gamma_{i}}{\lambda_{3}}, \quad
\omega_{i}=\frac{\beta_{i}(2u_{0}\lambda_{1}\lambda_{3}+\lambda_{1}\lambda_{4}-\lambda_{3}\lambda_{5})
+\gamma_{i}(2u_{0}\lambda_{2}\lambda_{3}+\lambda_{2}\lambda_{4}-\lambda_{3}\lambda_{6})}{4\lambda_{3}}.
\end{split}
\end{equation}

\textbf{Case~4.5:}
\begin{equation}\label{4.6}
\begin{split}
r_{2}&=-\frac{\lambda_{3}r_{1}+\lambda_{2}r_{3}}{\lambda_{1}}, \quad
r_{4}=\frac{r_{1}(\lambda_{3}\lambda_{5}-\lambda_{1}\lambda_{4}-2u_{0}\lambda_{1}\lambda_{3})
+r_{3}(\lambda_{2}\lambda_{5}-\lambda_{1}\lambda_{6})}{4\lambda_{1}}, \\
\alpha_{i}&=\frac{\beta_{i}(\lambda_{1}\lambda_{6}-\lambda_{2}\lambda_{5})-4\lambda_{2}\omega_{i}}
{2u_{0}\lambda_{2}\lambda_{3}+\lambda_{2}\lambda_{4}-\lambda_{3}\lambda_{6}}, \quad
\gamma_{i}=\frac{\beta_{i}(\lambda_{3}\lambda_{5}-\lambda_{1}\lambda_{4}-2u_{0}\lambda_{1}\lambda_{3})
+4\lambda_{3}\omega_{i}}{2u_{0}\lambda_{2}\lambda_{3}+\lambda_{2}\lambda_{4}-\lambda_{3}\lambda_{6}}.
\end{split}
\end{equation}

\textbf{Case~4.6:}
\begin{equation}\label{4.7}
\begin{split}
r_{3}&=-\frac{\lambda_{1}r_{2}+\lambda_{3}r_{1}}{\lambda_{2}}, \quad
r_{4}=\frac{r_{1}(\lambda_{3}\lambda_{6}-\lambda_{2}\lambda_{4}-2u_{0}\lambda_{2}\lambda_{3})
+r_{2}(\lambda_{1}\lambda_{6}-\lambda_{2}\lambda_{5})}{4\lambda_{2}}, \\
\alpha_{i}&=\frac{\gamma_{i}(\lambda_{2}\lambda_{5}-\lambda_{1}\lambda_{6})-4\lambda_{1}\omega_{i}}
{2u_{0}\lambda_{1}\lambda_{3}+\lambda_{1}\lambda_{4}-\lambda_{3}\lambda_{5}}, \quad
\beta_{i}=\frac{\gamma_{i}(\lambda_{3}\lambda_{6}-\lambda_{2}\lambda_{4}-2u_{0}\lambda_{2}\lambda_{3})
+4\lambda_{3}\omega_{i}}{2u_{0}\lambda_{1}\lambda_{3}+\lambda_{1}\lambda_{4}-\lambda_{3}\lambda_{5}}.
\end{split}
\end{equation}
Expressions \eqref{4.2} and \eqref{4.7} satisfy the constraint condition $\lambda_{2}(2u_{0}\lambda_{1}\lambda_{3}+\lambda_{1}\lambda_{4}-\lambda_{3}\lambda_{5})\neq0$. Expressions \eqref{4.3} and \eqref{4.6} satisfy the constraint condition $\lambda_{1}(2u_{0}\lambda_{2}\lambda_{3}+\lambda_{2}\lambda_{4}-\lambda_{3}\lambda_{6})\neq0$. Expressions \eqref{4.4} and \eqref{4.5} satisfy the constraint condition $\lambda_{3}(\lambda_{2}\lambda_{5}-\lambda_{1}\lambda_{6})\neq0$.

Combining expressions \eqref{4.5} and \eqref{4.1}, we get the interaction solutions of exponential functions and multiple mixed functions.
\begin{equation}\label{4.8}
\begin{split}
u_{D}&=2(\ln f_{D})_{xx}+u_{0}, \quad f_{D}=r_{0}\exp(\xi)+r_{5}\exp(-\xi)+\sum_{i=1}^{N}k_{i}\Lambda_{i}^{\rho_{i}}(\varphi_{i}),  \\
\xi&=r_{1}x+\frac{r_{1}(2u_{0}\lambda_{2}\lambda_{3}+\lambda_{2}\lambda_{4}-\lambda_{3}\lambda_{6})
+4\lambda_{2}r_{4}}{\lambda_{1}\lambda_{6}-\lambda_{2}\lambda_{5}}y
+\frac{r_{1}(2u_{0}\lambda_{1}\lambda_{3}+\lambda_{1}\lambda_{4}-\lambda_{3}\lambda_{5})
+4\lambda_{1}r_{4}}{\lambda_{2}\lambda_{5}-\lambda_{1}\lambda_{6}}z+r_{4}t, \\ \varphi_{i}&=-\frac{\lambda_{1}\beta_{i}+\lambda_{2}\gamma_{i}}{\lambda_{3}}x+\beta_{i}y+\gamma_{i}z
+\frac{\beta_{i}(2u_{0}\lambda_{1}\lambda_{3}+\lambda_{1}\lambda_{4}-\lambda_{3}\lambda_{5})
+\gamma_{i}(2u_{0}\lambda_{2}\lambda_{3}+\lambda_{2}\lambda_{4}-\lambda_{3}\lambda_{6})}{4\lambda_{3}}t.
\end{split}
\end{equation}

\begin{figure}[htb]
\centering
\includegraphics[width=0.325\textwidth]{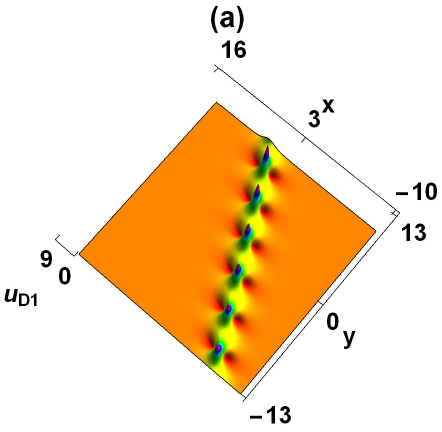}
\includegraphics[width=0.325\textwidth]{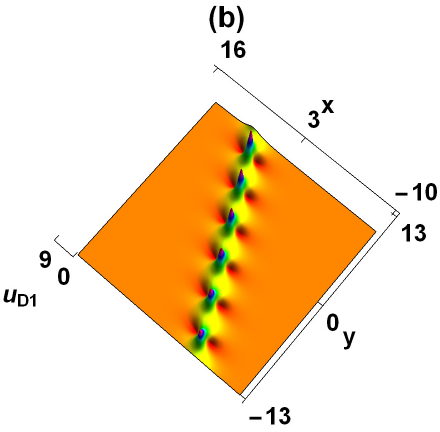}
\includegraphics[width=0.325\textwidth]{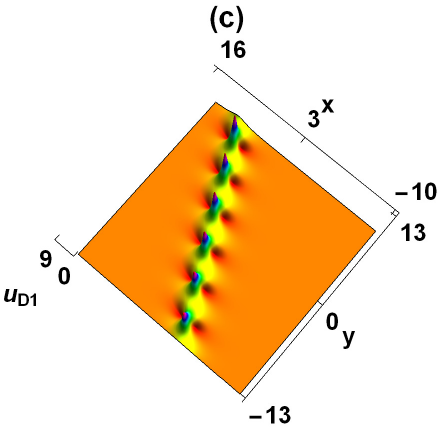}
\caption{\small Breather solution via the interaction solutions of exponential functions and multiple mixed functions \eqref{4.8} with $u_{0}=\lambda_{3}=\lambda_{5}=r_{4}=1,\lambda_{1}=3,
\lambda_{2}=\lambda_{4}=\lambda_{6}=k_{1}=2,r_{0}=1.5,r_{5}=1.2,r_{1}=-1,\beta_{1}=1.4,
\gamma_{1}=-2.2$. (a) $z=6,t=2,$ (b) $z=t=4,$ and (c) $z=2,t=6.$
}\label{fig:6}
\end{figure}

$N=\rho_{1}=1$ and $\Lambda_{1}=\cos$ are selected in the mixed solutions \eqref{4.8} to analyze the interaction phenomenon of different functions, we get some conclusions about the nonlinear effect. Fig.~\ref{fig:6} shows that one breather solution is composed of an upward hump and two downward valleys with periodicity on both sides of the horizontal plane.

We will take expressions \eqref{4.7} and \eqref{4.1} as an example to study the dynamics of the interaction solutions of exponential functions and multiple mixed functions.
\begin{equation}\label{4.9}
\begin{split}
u_{D}&=2(\ln f_{D})_{xx}+u_{0}, \quad f_{D}=r_{0}\exp(\xi)+r_{5}\exp(-\xi)+\sum_{i=1}^{N}k_{i}\Lambda_{i}^{\rho_{i}}(\varphi_{i}), \\
\xi&=r_{1}x+r_{2}y-\frac{\lambda_{1}r_{2}+\lambda_{3}r_{1}}{\lambda_{2}}z
+\frac{r_{1}(\lambda_{3}\lambda_{6}-\lambda_{2}\lambda_{4}-2u_{0}\lambda_{2}\lambda_{3})
+r_{2}(\lambda_{1}\lambda_{6}-\lambda_{2}\lambda_{5})}{4\lambda_{2}}t, \\
\varphi_{i}&=\frac{\gamma_{i}(\lambda_{2}\lambda_{5}-\lambda_{1}\lambda_{6})-4\lambda_{1}\omega_{i}}
{2u_{0}\lambda_{1}\lambda_{3}+\lambda_{1}\lambda_{4}-\lambda_{3}\lambda_{5}}x
+\frac{\gamma_{i}(\lambda_{3}\lambda_{6}-\lambda_{2}\lambda_{4}-2u_{0}\lambda_{2}\lambda_{3})
+4\lambda_{3}\omega_{i}}{2u_{0}\lambda_{1}\lambda_{3}+\lambda_{1}\lambda_{4}-\lambda_{3}\lambda_{5}}y
+\gamma_{i}z+\omega_{i}t.
\end{split}
\end{equation}

\begin{figure}[htb]
\centering
\includegraphics[width=0.325\textwidth]{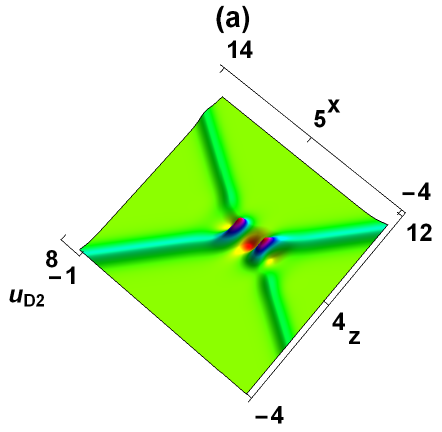}
\includegraphics[width=0.325\textwidth]{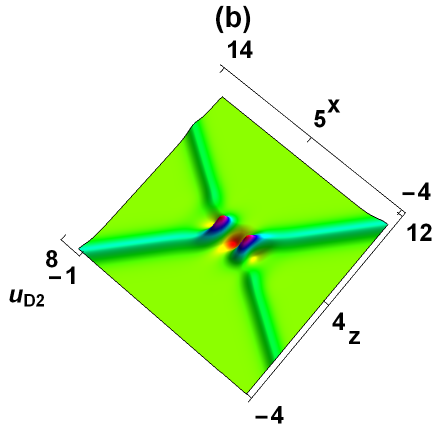}
\includegraphics[width=0.325\textwidth]{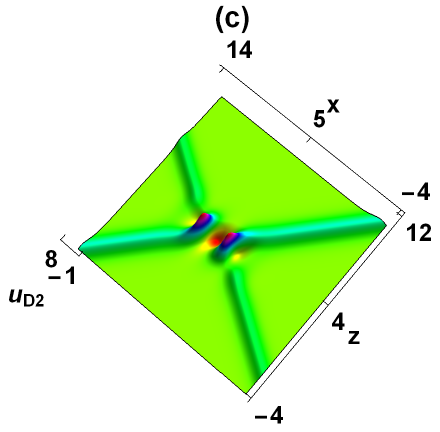}
\caption{\small Interactions between two kink waves and two rogue waves via the interaction solutions of exponential functions and multiple mixed functions \eqref{4.9} with $u_{0}=\lambda_{3}=\lambda_{5}=k_{1}=1$, $\lambda_{1}=k_{2}=3$, $\lambda_{2}=\lambda_{4}=\lambda_{6}=r_{0}=r_{5}=\omega_{2}=2$, $r_{1}=-0.2$, $r_{2}=1.5$, $\gamma_{1}=0.3$, $\omega_{1}=1.4$, $\gamma_{2}=-0.5$. (a) $y=6$, $t=2$, (b) $y=t=4$, and (c) $y=2$, $t=6$.
}\label{fig:7}
\end{figure}

$N=2$, $\rho_{1}=\rho_{2}=1$, $\Lambda_{1}=\cosh$ and $\Lambda_{2}=\cos$ are selected in the mixed solutions \eqref{4.9} to analyze the interaction phenomenon of different functions. Fig.~\ref{fig:7} shows that the interactions between two kink waves and two rogue waves, and two rogue waves are distributed at the intersection of the two kink waves. Two rogue waves collide with each other, resulting in two upward peaks and four downward valleys. It can be seen that the fusion between rogue waves leads to the change of their shape.

\subsection{Interaction solutions between products of different functions}

In order to obtain the interaction solutions between products of different functions can be given as follows:
\begin{equation}\label{4.10}
\begin{split}
f_{E}&=\sum_{j=1}^{M}s_{j}\Theta_{j}^{\tau_{j}}(\phi_{j})\Delta_{j}^{\sigma_{j}}(\delta_{j}), \quad
\phi_{j}=a_{j}x+b_{j}y+c_{j}z+d_{j}t, \quad \delta_{j}=m_{j}x+n_{j}y+p_{j}z+q_{j}t,
\end{split}
\end{equation}
where $s_{j}$, $a_{j}$, $b_{j}$, $c_{j}$, $d_{j}$, $m_{j}$, $n_{j}$, $p_{j}$, $q_{j}$ are all the real constants to be determined, $\tau_{j}$ and $\sigma_{j}$ are integers, $\Theta_{j}$ and $\Delta_{j}$ are any elementary functions. Substituting expressions \eqref{4.10} into Eq. \eqref{3.2}, we give the following results:

\textbf{Case~5.1:}
\begin{equation}\label{4.11}
\begin{split}
a_{j}&=\frac{b_{j}(\lambda_{1}\lambda_{6}-\lambda_{2}\lambda_{5})-4\lambda_{2}d_{j}}
{2u_{0}\lambda_{2}\lambda_{3}+\lambda_{2}\lambda_{4}-\lambda_{3}\lambda_{6}}, \quad
c_{j}=\frac{b_{j}(\lambda_{3}\lambda_{5}-\lambda_{1}\lambda_{4}-2u_{0}\lambda_{1}\lambda_{3})
+4\lambda_{3}d_{j}}{2u_{0}\lambda_{2}\lambda_{3}+\lambda_{2}\lambda_{4}-\lambda_{3}\lambda_{6}}, \\
p_{j}&=-\frac{\lambda_{1}n_{j}+\lambda_{3}m_{j}}{\lambda_{2}}, \quad
q_{j}=\frac{m_{j}(\lambda_{3}\lambda_{6}-\lambda_{2}\lambda_{4}-2u_{0}\lambda_{2}\lambda_{3})
+n_{j}(\lambda_{1}\lambda_{6}-\lambda_{2}\lambda_{5})}{4\lambda_{2}},
\end{split}
\end{equation}
where $\lambda_{2}(2u_{0}\lambda_{2}\lambda_{3}+\lambda_{2}\lambda_{4}-\lambda_{3}\lambda_{6})\neq0$.

\textbf{Case~5.2:}
\begin{equation}\label{4.12}
\begin{split}
b_{j}&=-\frac{\lambda_{3}a_{j}+\lambda_{2}c_{j}}{\lambda_{1}}, \quad
d_{j}=\frac{a_{j}(\lambda_{3}\lambda_{5}-\lambda_{1}\lambda_{4}-2u_{0}\lambda_{1}\lambda_{3})
+c_{j}(\lambda_{2}\lambda_{5}-\lambda_{1}\lambda_{6})}{4\lambda_{1}}, \\
m_{j}&=\frac{p_{j}(\lambda_{2}\lambda_{5}-\lambda_{1}\lambda_{6})-4\lambda_{1}q_{j}}
{2u_{0}\lambda_{1}\lambda_{3}+\lambda_{1}\lambda_{4}-\lambda_{3}\lambda_{5}}, \quad
n_{j}=\frac{p_{j}(\lambda_{3}\lambda_{6}-\lambda_{2}\lambda_{4}-2u_{0}\lambda_{2}\lambda_{3})
+4\lambda_{3}q_{j}}{2u_{0}\lambda_{1}\lambda_{3}+\lambda_{1}\lambda_{4}-\lambda_{3}\lambda_{5}},
\end{split}
\end{equation}
where $\lambda_{1}(2u_{0}\lambda_{1}\lambda_{3}+\lambda_{1}\lambda_{4}-\lambda_{3}\lambda_{5})\neq0$.

\textbf{Case~5.3:}
\begin{equation}\label{4.13}
\begin{split}
a_{j}&=-\frac{\lambda_{1}b_{j}+\lambda_{2}c_{j}}{\lambda_{3}}, \quad
d_{j}=\frac{b_{j}(2u_{0}\lambda_{1}\lambda_{3}+\lambda_{1}\lambda_{4}-\lambda_{3}\lambda_{5})
+c_{j}(2u_{0}\lambda_{2}\lambda_{3}+\lambda_{2}\lambda_{4}-\lambda_{3}\lambda_{6})}{4\lambda_{3}}, \\
p_{j}&=\frac{n_{j}(\lambda_{3}\lambda_{5}-\lambda_{1}\lambda_{4}-2u_{0}\lambda_{1}\lambda_{3})
+4\lambda_{3}q_{j}}{2u_{0}\lambda_{2}\lambda_{3}+\lambda_{2}\lambda_{4}-\lambda_{3}\lambda_{6}}, \quad
m_{j}=\frac{n_{j}(\lambda_{1}\lambda_{6}-\lambda_{2}\lambda_{5})-4\lambda_{2}q_{j}}
{2u_{0}\lambda_{2}\lambda_{3}+\lambda_{2}\lambda_{4}-\lambda_{3}\lambda_{6}},
\end{split}
\end{equation}
where $\lambda_{3}(2u_{0}\lambda_{2}\lambda_{3}+\lambda_{2}\lambda_{4}-\lambda_{3}\lambda_{6})\neq0$.

\textbf{Case~5.4:}
\begin{equation}\label{4.14}
\begin{split}
n_{j}&=-\frac{\lambda_{3}m_{j}+\lambda_{2}p_{j}}{\lambda_{1}}, \quad
q_{j}=\frac{m_{j}(\lambda_{3}\lambda_{5}-\lambda_{1}\lambda_{4}-2u_{0}\lambda_{1}\lambda_{3})
+p_{j}(\lambda_{2}\lambda_{5}-\lambda_{1}\lambda_{6})}{4\lambda_{1}},  \\
b_{j}&=\frac{a_{j}(2u_{0}\lambda_{2}\lambda_{3}+\lambda_{2}\lambda_{4}-\lambda_{3}\lambda_{6})
+4\lambda_{2}d_{j}}{\lambda_{1}\lambda_{6}-\lambda_{2}\lambda_{5}}, \quad
c_{j}=\frac{a_{j}(2u_{0}\lambda_{1}\lambda_{3}+\lambda_{1}\lambda_{4}-\lambda_{3}\lambda_{5})
+4\lambda_{1}d_{j}}{\lambda_{2}\lambda_{5}-\lambda_{1}\lambda_{6}},
\end{split}
\end{equation}
where $\lambda_{1}(\lambda_{1}\lambda_{6}-\lambda_{2}\lambda_{5})\neq0$.

\textbf{Case~5.5:}
\begin{equation}\label{4.15}
\begin{split}
a_{j}&=\frac{c_{j}(\lambda_{2}\lambda_{5}-\lambda_{1}\lambda_{6})-4\lambda_{1}d_{j}}
{2u_{0}\lambda_{1}\lambda_{3}+\lambda_{1}\lambda_{4}-\lambda_{3}\lambda_{5}}, \quad
b_{j}=\frac{c_{j}(\lambda_{3}\lambda_{6}-\lambda_{2}\lambda_{4}-2u_{0}\lambda_{2}\lambda_{3})
+4\lambda_{3}d_{j}}{2u_{0}\lambda_{1}\lambda_{3}+\lambda_{1}\lambda_{4}-\lambda_{3}\lambda_{5}}, \\
n_{j}&=\frac{m_{j}(2u_{0}\lambda_{2}\lambda_{3}+\lambda_{2}\lambda_{4}-\lambda_{3}\lambda_{6})
+4\lambda_{2}q_{j}}{\lambda_{1}\lambda_{6}-\lambda_{2}\lambda_{5}}, \quad
p_{j}=\frac{m_{j}(2u_{0}\lambda_{1}\lambda_{3}+\lambda_{1}\lambda_{4}-\lambda_{3}\lambda_{5})
+4\lambda_{1}q_{j}}{\lambda_{2}\lambda_{5}-\lambda_{1}\lambda_{6}},
\end{split}
\end{equation}
where $(\lambda_{1}\lambda_{6}-\lambda_{2}\lambda_{5})
(2u_{0}\lambda_{1}\lambda_{3}+\lambda_{1}\lambda_{4}-\lambda_{3}\lambda_{5})\neq0$.

\textbf{Case~5.6:}
\begin{equation}\label{4.16}
\begin{split}
c_{j}&=-\frac{\lambda_{1}b_{j}+\lambda_{3}a_{j}}{\lambda_{2}}, \quad
d_{j}=\frac{a_{j}(\lambda_{3}\lambda_{6}-\lambda_{2}\lambda_{4}-2u_{0}\lambda_{2}\lambda_{3})
+b_{j}(\lambda_{1}\lambda_{6}-\lambda_{2}\lambda_{5})}{4\lambda_{2}},  \\
m_{j}&=-\frac{\lambda_{1}n_{j}+\lambda_{2}p_{j}}{\lambda_{3}}, \quad
q_{j}=\frac{n_{j}(2u_{0}\lambda_{1}\lambda_{3}+\lambda_{1}\lambda_{4}-\lambda_{3}\lambda_{5})
+p_{j}(2u_{0}\lambda_{2}\lambda_{3}+\lambda_{2}\lambda_{4}-\lambda_{3}\lambda_{6})}{4\lambda_{3}},
\end{split}
\end{equation}
where $\lambda_{2}\lambda_{3}\neq0$.

Substituting relational formula \eqref{4.10} and the expression \eqref{4.13} into transformation \eqref{3.1}, corresponding interaction solutions between products of different functions for Eq. \eqref{1.1} appear as
\begin{equation}\label{4.17}
\begin{split}
u_{E}&=2(\ln f_{E})_{xx}+u_{0}, \quad f_{E}=\sum_{j=1}^{M}s_{j}\Theta_{j}^{\tau_{j}}(\phi_{j})\Delta_{j}^{\sigma_{j}}(\delta_{j}), \\
\phi_{j}&=-\frac{\lambda_{1}b_{j}+\lambda_{2}c_{j}}{\lambda_{3}}x+b_{j}y+c_{j}z
+\frac{b_{j}(2u_{0}\lambda_{1}\lambda_{3}+\lambda_{1}\lambda_{4}-\lambda_{3}\lambda_{5})
+c_{j}(2u_{0}\lambda_{2}\lambda_{3}+\lambda_{2}\lambda_{4}-\lambda_{3}\lambda_{6})}{4\lambda_{3}}t, \\ \delta_{j}&=\frac{n_{j}(\lambda_{1}\lambda_{6}-\lambda_{2}\lambda_{5})-4\lambda_{2}q_{j}}
{2u_{0}\lambda_{2}\lambda_{3}+\lambda_{2}\lambda_{4}-\lambda_{3}\lambda_{6}}x+n_{j}y
+\frac{n_{j}(\lambda_{3}\lambda_{5}-\lambda_{1}\lambda_{4}-2u_{0}\lambda_{1}\lambda_{3})
+4\lambda_{3}q_{j}}{2u_{0}\lambda_{2}\lambda_{3}+\lambda_{2}\lambda_{4}-\lambda_{3}\lambda_{6}}z+q_{j}t.
\end{split}
\end{equation}

\begin{figure}[htb]
\centering
\includegraphics[width=0.325\textwidth]{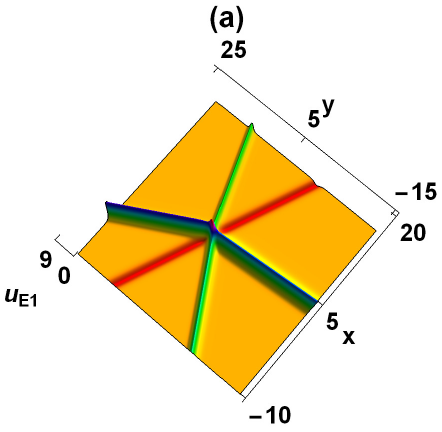}
\includegraphics[width=0.325\textwidth]{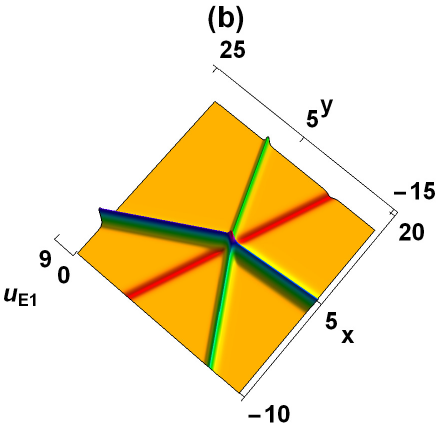}
\includegraphics[width=0.325\textwidth]{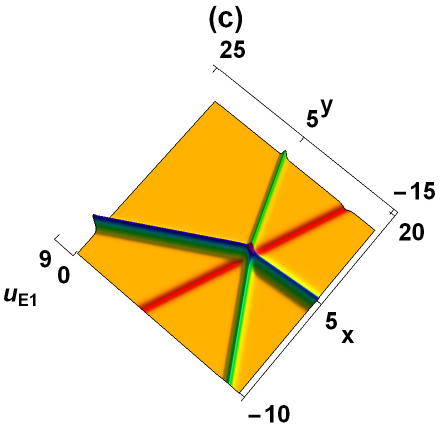}
\caption{\small Interactions between three bell shaped waves via the interaction solutions between products of different functions \eqref{4.17} with $u_{0}=\lambda_{3}=\lambda_{5}=1$, $\lambda_{1}=3$, $\lambda_{2}=\lambda_{4}=\lambda_{6}=2$, $s_{1}=1.2$, $s_{2}=1.5$, $b_{1}=-1$, $c_{1}=1.4$, $n_{1}=0.6$, $q_{1}=0.2$, $b_{2}=-1.4$, $c_{2}=0.75$, $n_{2}=-1.2$, $q_{2}=-0.1$. (a) $z=6$, $t=2$, (b) $z=t=4$, and (c) $z=2$, $t=6$.
}\label{fig:8}
\end{figure}

$M=2$, $\tau_{1}=\tau_{2}=\sigma_{1}=\sigma_{2}=1$, $\Theta_{1}=\Theta_{2}=\cosh$, $\Delta_{1}=\exp$ and $\Delta_{2}=\sech$ are selected in the interaction solutions between products of different functions \eqref{4.17} to analyze collisions between multiple waves. Fig.~\ref{fig:8} analyzes the mutual collision between three bell shaped waves, forming two upward bell shaped waves and one downward bell shaped waves. It can be seen from Fig.~\ref{fig:8} that the downward bell shaped wave is linear and does not change, but the two upward bell shaped waves change due to collision. Two upward bell shaped waves tilt in the opposite direction respectively, and a peak soliton is generated at the intersection of the three bell shaped waves.

\section{Conclusion}
\label{s:Conclusion}

In this paper, the integrability of (3+1)-dimensional KdV-CBS equation is studied by means of Bell polynomial theory, and it is found that it is completely integrable in Lax pair sense. Then the B\"{a}cklund transformation, Lax pair and infinite conservation laws of the equation are constructed. Some new formal solutions of Eq. \eqref{1.1} are studied by the Homoclinic test method and symbolic computations, including the lump solutions, multiple mixed function solutions, hybrid solutions composed of lump solitons and multiple functions, interaction solutions of exponential functions and multiple mixed functions, interaction solutions between products of different functions. We graphically illustrate the obtained results with specific parameters. The successfully derived mixed wave solutions are visually represented in the images, which proves the effectiveness of the method and provides important results for analysis and practical application. The corresponding work extends the types of mixed solutions, these analytical solutions can provide shallow water wave models for some nonlinear phenomena in physics. These solutions are important because they have the ability to improve our knowledge of wave behavior and provide insightful information to disciplines like tsunami modeling and coastal engineering.

\section*{CRediT authorship contribution statement}
\textbf{Peng-Fei Han:} Conceptualization, Formal analysis, Investigation, Methodology, Validation, Writing--original draft, Writing--review $\&$ editing. \textbf{Yi Zhang:} Conceptualization, Supervision, Validation, Writing--review $\&$ editing.

\section*{Acknowledgement}
This work was supported by the National Natural Science Foundation of China (Grant Nos. 11371326, 11975145 and 12271488).

\section*{Conflict of interests}
The authors declare that there is no conflict of interests regarding the research effort and the publication of this paper.

\section*{Data availability statements}
All data generated or analyzed during this study are included in this published article.

\section*{ORCID}
{\setlength{\parindent}{0cm}
Peng-Fei Han: https://orcid.org/0000-0003-1164-5819 \\
Yi Zhang: https://orcid.org/0000-0002-8483-4349}

\end{document}